\documentclass{article}
\font\cero=cmss10 scaled 1728 \font\uno=cmssbx10 scaled 1200
\usepackage{tensor}
\usepackage{verbatim}
\usepackage{amsfonts}
\usepackage{graphicx}
\usepackage{float}
\begin{document}
\begin{flushleft}
{\cero Deforming the theory $\lambda\phi^{4}$ along the parameters and fields gradient flows}\\
\end{flushleft}
{\sf R. Cartas-Fuentevilla}\\
{\it Instituto de F\'{\i}sica, Universidad Aut\'onoma de Puebla,
Apartado postal J-48 72570 Puebla Pue., M\'exico}; cartas@sirio.ifuap.buap.mx/TEL-FAX (01222) 2295611

\noindent{\sf A. Olvera-Santamaria} \\
{\it Facultad de Ciencias F\'{\i}sico-Matem\'{a}ticas, Universidad Aut\'onoma de Puebla,
Apartado postal 1152, 72001  Puebla Pue., M\'exico.} \\

Considering the action for the theory $\lambda\phi^{4}$ for a massive scalar bosonic field as an entropy functional on the space of coupling constants and on the space of fields, we determine the gradient flows for the scalar field, the mass, and the self-interaction parameter. When the flow parameter is identified with the energy scale,  we show that there exist phase transitions  between unbroken exact symmetry scenarios and spontaneous symmetry breaking scenarios at increasingly high energies. Since a non-linear heat equation drives the scalar field through a {\it reaction-diffusion} process, in general the flows are not reversible, mimicking the renormalization group flows of the $c$-theorem; the deformation of the field at increasingly high energies can be described as non-linear traveling waves, or solitons associated to self-similar solutions.\\



\noindent {\uno I. Introduction} \vspace{1em}

The essential idea behind the Renormalization group flow (RGF) for a quantum field theory is that the physics depends on the scale; hence, the description of a system in terms of both the physical degrees of freedom, and an action principle will change with the scale. The usual approach for RGF consist of determining the differential equations that govern the behavior of the corresponding coupling constants under changes of scale; the $\beta$-functions perturbatively evaluated in a loop expansion, determine the form of those equations, defining a flow on the space of coupling constants. The problem whether such a flow is a gradient flow is a very intricate question from physical and mathematical point of view; the problem consist of determining the orthogonality of that flow to level surfaces of a potential function on the space of coupling constants.

On the other hand, geometrical evolution equations were originally introduced to address the problem of uniformization theorems in differential geometry, and successfully implemented by G. Perelman by probing in the positive the Poincar\'e conjecture in three-dimensions \cite{perelman}; the monotonic character of the so called {\it entropy functional} plays the key role in the construction of the corresponding gradient flow, leading to a scenario similar to the problem gradient flow/potential-function in the context of the renormalization group flows mentioned in the previous paragraph, and the possible relationship between both scenarios is not understood nowadays. Furthermore, the entropy functionals in the context of the geometrical flows appear in physical contexts playing the role of action principles, and it is well known the appearance of that used by G. Perelman in the context of string theory,  since the celebrated Ricci flow corresponds to the lowest order term in the perturbation loop expansion; hence, in principle, the geometrical flows can be used as approximations to RGF in quantum field theory, since the former are naturally embedded into the later.

The Perelman entropy functional approach for geometrical flows suggests a direct recipe for determining relevant flows for a theory of interest, and consist of taking the corresponding action as a potential function on the space of coupling constant/field space along a flow parameter (which can be identified in particular with the energy scale, for instance), and invoking monotonicity, to determine the corresponding gradient flows of its coupling constants, mass parameters, and fields appearing in such an action; such gradient flows can be compared subsequently with the standard RGF defined by the $\beta$-functions. Independently on the similarities and differences between both flows, the gradient flows of an action have interest by themselves; such an approach can be considered an a non-perturbative approximation to the RGF in the sense that the gradient flows are not constructed in terms of the $\beta$-functions determined order by order in loop expansion. However, as we shall see, gradient flows as well will be defined in terms of infinite sums. 

As already commented, the largely motivation in the study of geometric flows is the topological classification of manifolds, and thus the geometrical evolution equations describe the dynamics of a background manifold; however, in the present work we attempt to apply the machinery of geometrical flows to the study of the simplest interacting field theory, that of one-component scalar bosonic field with quartic self-interaction $\lambda\phi^{4}$ in a background that,  as usual in quantum field theory, is flat and fixed; with the appropriate restrictions the theory is bounded from below, and offers  spontaneous symmetry breaking scenarios, which will be studied dynamically along the flows. This theory, although simple is not trivial, even in four-dimensions; the possible triviality of the theory is intimately related with the strongest theoretical upper bounds on the values of the Higgs mass \cite{podo}. However, our treatment has no 
restrictions on the dimension of the background, and will be developed for $\lambda\phi^{4}_{d}$, with $d$ arbitrary.

In the next section we define the Lagrangian for the theory where the scalar field $\phi$ and parameters ($m$, $\lambda$), depend on a flow parameter $\tau$; the $\tau$-gradient of the Lagrangian is determined as a function of the gradient flows of the field and of the parameters. Considering then the steepest descent and ascent for that Lagrangian  (in this sense it works as a entropy functional on both the space of fields and the space of parameters), we determine those gradient flows. The steepest descent (Section III) and the steepest ascent (Section IV), are analyzed separately;  in both cases the gradient flows 
for the parameters $(m, \lambda)$ can be solved explicitly; under the restriction of a stationary scalar field, the potential of the theory shows a qualitative change  between Unbroken Exact Symmetry (UES) scenarios and Spontaneous Symmetry Breaking (SSB) scenarios along of certain trajectories; in particular, the change of a unique vacuum in the UES scenarios to the degenerate vacua in the SSB scenarios (and vice versa) is achieved continuously at finite {\it time}.  Along other trajectories there no exists a qualitative change in the vacuum structure. In Section V, we show that the parameters flows correspond in general to infinite sums, mimicking the loop expansion for the $\beta$ functions in the usual RG  parameters evolution; the phase transitions between UES and SSB scenarios are described in this section. In Section VI, we consider the phase transitions generated by parameters gradient flows {\it in phase}, and {\it out of phase}. In Section VII we switch off the parameters flows, and switch on the scalar field flow through a {\it reaction-diffussive} evolution; considering the case of one spatial dimension, explicit solutions for for the non-linear evolution are discussed, namely, {\it traveling waves} solutions, and {\it solitons as self-similar} solutions. We conclude with possible extensions, and remarks.

It is important to clarify in that sense the concept of {\it duality} will be used in this work; nowadays duality appears linking strong and weak coupling regions in string theory and supersymmetric field theories; additionally it links electric phase/magnetic phase, Higgs phase/confinement, etc. Since we shall deform the theory along the trajectories in the space of parameters and fields,  the different descriptions of the theory will emerge naturally in the regions of strong/weak self-interactions, massive/massless phases, etc.; similarly in the case of the different phases described by the deformed dynamical scalar field $\phi$. Hence, the duality relations used here, are closer in spirit to the nonperturbative S duality, as opposed to the T duality that is perturbative in nature. Furthermore, we use the Ehrenfest classification for phase transitions of a system; a phase transition is characterized by the order given by the lowest derivative of the energy that is not continuous at the transition point. Since in our case, the scalar field is dynamically deformed along the $\tau$-gradient, the functional whose behavior determines the phase transition must include also the kinetic part, besides the usual term associated to the potential energy. In fact, all phase transitions that occur in the present analysis will be of first order. \\

\noindent {\uno II.  The Lagrangian of the theory $\lambda\phi^{4}_{d}$ as an entropy functional} 
\vspace{1em}

We start with the usual Lagrangian for the theory $\lambda\phi^{4}_{d}$, upgrading it to a functional of a parameter $\tau$ that defines a flow in the space of parameters and fields;
\begin{equation}
{\cal E}[\phi(x,\tau), m(\tau), \lambda(\tau)] = \int dx^{d}(\frac{1}{2}\partial^{i}\phi\cdot\partial_{i}\phi-\frac{1}{2}m^2\phi^2-\frac{1}{4!}\lambda\phi^4);
\label{lag}
\end{equation}
the $\tau$-dependence of the functional (\ref{lag}) can be interpreted in various senses; if $\tau$ is identified with the energy scale, then the present work will allow, along the Perelman-Hamilton scheme, study the deformation of the theory conform to $\tau$-variations. Furthermore, $\tau$ can also be identified with the temperature, and since the energy of the system will change with $\tau$, then ${\cal E}={\cal E}(\tau)$ can describe an open system, exchanging energy with a surrounding thermal reservoir; hence, we have a field theory at non-zero temperature, with ${\cal E}(\tau=0)$ identified with the usual Lagrangian for a theory {	\it at zero-temperature}.

In (\ref{lag}), the coupling constant $\lambda$ is usually assumed to be positive; the unbroken exact symmetry scenario requires 
$m^2>0$, and the spontaneous symmetry breaking requires $m^2<0$. These restrictions on the mass define regimes in the space of parameters, which will be connected by trajectories defined along the gradient flows; hence, such trajectories will connect UES  and SSB scenarios. 
The gradient along the flow parameter $\tau$ reads
\begin{equation}
     \partial_{\tau}{\cal E} = -\int dx^{d}  [( {\Box}\phi +m^2\phi+\frac{1}{3!}\lambda\phi^3) \partial_{\tau}\phi + \frac{1}{2}\phi^{2} \frac{d}{d \tau}m^{2} + \frac{1}{4!}\phi^{4} \frac{d}{d \tau}\lambda]; \label{var-2}
\end{equation}
therefore, the following flows determine the steepest descent ($a=1$) and the steepest ascent ($a=-1$) for ${\cal E}$:
\begin{eqnarray}
     a\frac{d}{d \tau}m^{2} \!\! & = & \!\! \mu (m^{2})^{n}; \qquad \mu \geq 0, \qquad n = 0, \pm 2, \pm4, \ldots \label{gradient1} \\
     a\frac{d}{d \tau} \lambda \!\! & = & \!\! \nu \lambda^{l}; \qquad \nu \geq 0, \qquad l = 0, \pm 2, \pm4, \ldots \label{gradient2} \\
     a\partial_{\tau} \phi \!\! & = & \!\! \Box\phi + m^{2}\phi + \frac{\lambda}{3!} \phi^{3}, \label{gradient3}
\end{eqnarray} 
with $\mu$ and $\nu$ are positive constants; under these flows we have
\begin{equation}
     \partial_{\tau} {\cal E} = - a \int d x^{d} [(\Box\phi + m^{2}\phi + \frac{1}{3!} \lambda \phi^{3})^{2} + \frac{\mu}{2} \phi^{2} (m^{2})^{n} + \frac{\nu}{4!} \phi^{4} \lambda^{l}],
     \label{pn}
\end{equation} 
which is strictly negative ($a=1$), or strictly positive ($a=-1$). Note that, at the level of the gradients (\ref{gradient1})--(\ref{pn}), the case $a=1$ represents in relation to the case $a=1$, the backward evolution obtained by the reversion $\tau \rightarrow -\tau$; due to the presence of a {\it heat}-like equation in (\ref{gradient3}), the evolution is not reversible. These flows are defined modulo reparametrizations of $\tau$, which will be used when we shall face the appearance of singularities along the flows. Therefore, when such singularities are not removed, then we shall have a phase transition.

For both cases $a=\pm 1$, the stationary points of the non-linear {\it heat} equation (\ref{gradient3}), are given by the space of solutions of the classical equations of motion, $\Box\phi + m^{2}\phi + \frac{\lambda}{3!}\phi^{3} =0$; thus, the flow (\ref{gradient3}) can be understood as a diffusive process acting on the field $\phi$, deforming the field configurations given by the usual classical solutions. Thus, in the present scheme, the evolution of the parameters is accompanied by the diffusion of the dynamical field $\phi$, which does not appear explicitly in Eqs. (\ref{gradient1}), and (\ref{gradient2}); however, the parameters as functions on $\tau$ appear in the diffusion equation (\ref{gradient3}).

The behavior of the flows depends sensitively on the value of $a$, hence, we develop each case separately. By simplicity we consider first the deformation of the theory around the stationary con\-fi\-gurations for the field $\phi$, {\it i.e.} around the classical solutions space of the theory; therefore, the kinetic part of the theory is maintained stationary. In Section VII, we face the diffusion of the field $\phi$ and its impact in the deformation of the theory along the flows.\\

\noindent {\uno III. The steepest descent: $a=1$}
\vspace{1em}

The solutions for the equations (\ref{gradient1}), and (\ref{gradient2}), can be found directly,
\begin{eqnarray}
     m^{2} (n,\tau) \!\! & = & \!\! [(n-1) (-\mu\tau + \mu_{0})]^{\frac{1}{1-n}}, \label{mass}\\
     \lambda (l,\tau) \!\! & = & \!\! [(l-1) (-\nu\tau + \nu_{0})]^{\frac{1}{1-l}}, \label{lambda}
\end{eqnarray}
where $\mu_{0}$, and $\nu_{0}$, are integration constants, without positivity restrictions, as opposed to $\mu$, and $\nu$;   
since $\lambda$ is assumed to be positive from the beginning, a positive gradient (\ref{gradient2}) will maintain positivity for all $\tau$, and 
in general the behavior of $\lambda$ does not modify qualitatively  the deformation of the theory. We comment timely on the general case with $\lambda$ variable; in the Section IV corresponding to the steepest ascent case, we analyze the limit $\lambda \rightarrow 0$, in which the theory is no longer self-interacting, and becomes free.
Under these conditions the deformation of the potential $V= \frac{1}{2}m^{2}(\tau)\phi^{2}(x)+ \frac{1}{4!}\lambda\phi^{4}(x)$, depends essentially on the behavior of the mass.  

\begin{figure}[H]
  \begin{center}
    \includegraphics[width=.6\textwidth]{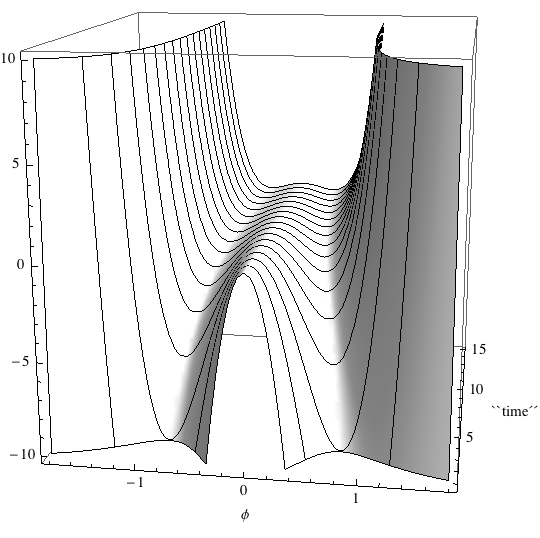}
  \caption{}  
   \label{fig1}
  \end{center}
\end{figure}

Let us consider that the initial point of the trajectories is at  $\tau=0$; hence $m^{2}(n, \tau =0)= [(n-1) \mu_{0}]^{\frac{1}{1-n}}$, and the sign of $m^{2}$ depends on the possible values of $n$ and the sign of $\mu_{0}$, and in this manner, the following cases are in order:

{\uno i)} $\mu_{0}<0$, and  $n\geq 2$, with $m^2(n, \tau =0)<0$: In this case we have a SSB scenario at $\tau=0$, which is stable at increasingly high energy; the deformation of the potential is shown in the figure \ref{fig1}. 
In this case, the mass, the Lagrangian (\ref{lag}), and the gradient (\ref{pn}) are well behaved along the flow. 
the condition $m^2<0$ for a SSB scenario, is maintained stable in the interval $\tau \in [0,+\infty)$; the two lowest-energy states are localized at    $\phi_{vacuum}(n,\tau)= \pm 2\sqrt{-\frac{m^2(n,\tau)}{\lambda}}$, and with a height given by $V_{vacuum}= -\frac{m^4}{\lambda}$; according to the Eq.(\ref{mass}), we have that $lim_{\tau\rightarrow +\infty}m^2(n,\tau)=0$, and then $lim_{\tau\rightarrow +\infty}\phi_{vacuum}(n,\tau)=0=lim_{\tau\rightarrow +\infty}V_{vacuum}(n,\tau)$, and then the evolution to higher energies tends to eliminate the degeneration, with a potential that is soften as $\tau$ is increasing; additionally we have that $lim_{n\rightarrow+\infty}m^2(n,\tau)=m^2(n=2,\tau)$. In this case the evolution is smooth, without singularities or critical values for $n$, or $\tau$.

In the case of a parameter $\lambda$ variable, the condition $\lambda(l,\tau=0)>0$, is preserved under a strict positive gradient, with $lim_{\tau\rightarrow +\infty}\lambda(l,\tau)=+\infty$; therefore, the deformation of the potential shown in the figure 1 is valid, but such a deformation is {\it faster} as $\tau\rightarrow +\infty$. Conversely, in the case of a stationary mass, the effect of a gradient-flow for  $\lambda$  is exactly that shown in the figure 1. Hence, the self-coupling parameter become increasingly large, an effect contrary to that of the known
{\it asymptotic freedom} of gauge theory; such an effect will be present in the case $a=-1$ discussed below.

Therefore, considering the ``long-$\tau$" behavior $lim_{\tau\rightarrow +\infty}(m^2,\lambda)=(0,+\infty)$, the theory becomes asymptotically  massless and strongly self-interacting; complementarily, at the ``low-$\tau$" region, the theory will become massive and weakly self-interacting. These {\it dual} versions of the theory are connected smoothly by a trajectory in the space of parameters.

{\uno ii)} $\mu_{0}<0$, and  $n\leq 0$, with $m^2(n, \tau =0)>0$: In this case we have an UES scenario at $\tau=0$, and according to Eq. (\ref{mass}), is stable in the interval  $\tau \in [0,+\infty)$. The asymptotic limits for the mass are $lim_{\tau\rightarrow +\infty}m^2(n,\tau)=+\infty$, which  is not bounded as $\tau$ is increasing, as opposed to the previous case; additionally we have that  $lim_{n\rightarrow-\infty}m^2(n,\tau)=m^2(n=0,\tau)$.  Such as the previous case, the mass, the Lagrangian (\ref{lag}) and the gradient (\ref{pn}) are well behaved along the flow.

If the potential undergoes additionally a positive gradient for $\lambda$, then similarly to the previous case, the deformation is {\it faster} in relation to that caused only by the mass flow. In the case of a stationary mass, then the gradient flow for $\lambda$ will cause essentially the same effect on the potential described in the previous paragraph.
In this case we have a massive and strongly self-interacting theory in the high-energy region, and  a massless and weakly self-interacting {\it dual} theory in the low-energy region.

{\uno iii)} $\mu_{0}>0$, and  $n\geq 2$, with $m^2(n, \tau =0)>0$: This case represents in relation to the case i), only a change of sign in the constant $\mu_{0}$, and respect to to case ii) a simultaneous change of sign for $\mu_{0}$ and $n$;
 thus, we have an UES at $\tau=0$;  however, as opposed to the case ii), this scenario is not stable for all $\tau$-domain (for this particular parametrization), since at $\tau=\frac{\mu_{0}}{\mu}$, there exists a singularity in the mass evolution. Therefore, the UES scenario is stable only in the interval $\tau\in[0,\frac{\mu_{0}}{\mu})$, and on the other side of the singularity the mass has changed of sign, and hence we have a SSB scenario in the complementary domain $\tau\in (\frac{\mu_{0}}{\mu}, +\infty)$, with an asymptotic limit similar to that of the case i),  $lim_{\tau\rightarrow +\infty}m^2(n,\tau)=0$. Therefore, a qualitative change from an one lowest-energy state(UES) to two lowest-energy states  (SSB) is achieved in finite `time";  the figure 2 shows 
 this (non-continuous) qualitative change. At the discontinuity we have $lim_{\tau\rightarrow _{\frac{\mu_{0}}{\mu}}}m^2(n,\tau)=\pm\infty$, and adidtionally the gradient (\ref{pn}) is discontinuous at the transition. Thus, one may be tempted to identify this discontinuity as a phase transition.  
\begin{figure}[H]
  \begin{center}
    \includegraphics[width=.65\textwidth]{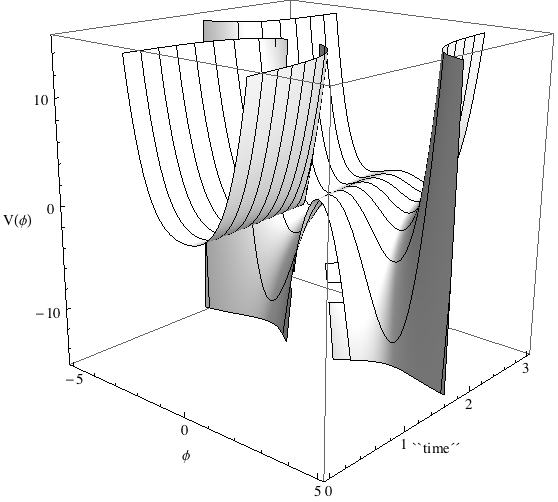}
  \caption{}  
   \label{fig:ejemplo}
  \end{center}
\end{figure}
However, this qualitative change is not physically meaningful, since a $\tau$-reparametrization allows to remove such a discontinuity. Consider for example, the reparametrization of the gradient Eq. (\ref{gradient1}) (positive for $a=1$), given by $\frac{d}{d\tau}m^2= \gamma^{2} e^{-\gamma\tau}(m^2)^n$, with $\gamma$
a  constant, and we have again $n\geq 2$; the solution is given by $m^2(n,\tau)= [(n-1)\gamma e^{-\gamma\tau}]^{\frac{1}{1-n}}$. Therefore $m^2(n,\tau=0)= [(n-1)\gamma ]^{\frac{1}{1-n}}$, and thus we need to restrict $\gamma>0$, in order to have a UES scenario at $\tau=0$ with $m^2(n, \tau =0)>0$. This solution does not show singularities at the complete interval $\tau\in[0,+\infty)$, as wanted. Additionally we have that $lim_{\tau\rightarrow +\infty}m^2(n,\tau)=+\infty$, and the UES scenario is stable; the effect of a gradient flow for $\lambda$ is similar to the previous case ii). This case also is similar to the case ii) in relation to the {\it dual} versions of the theory in the high and low energy regions.

{\uno iv)} $\mu_{0}>0$, and  $n\leq 0$, with $m^2(n, \tau =0)<0$: this case corresponds then to an SSB scenario at $\tau=0$; however, the negativity of $m^2$
is maintained only in the finite interval $\tau\in[0,\frac{\mu_{0}}{\mu})$, since $m^2(n, \tau=\frac{\mu_{0}}{\mu})=0$. Furthermore, in the interval $\tau\in[\frac{\mu_{0}}{\mu},+\infty)$ we have that $m^2>0$, and then an UES scenario; therefore, the mass evolves continuously from a SSB scenario to  an UES scenario, 
crossing at finite ``time" the massless scenario at $\tau=\frac{\mu_{0}}{\mu}$, and with $lim_{\tau\rightarrow +\infty}m^2(n,\tau)=+\infty$. This evolution is shown in figure 3.

However, although the mass is well behaved for all values of $n$, there exists a discontinuity with the gradient (\ref{pn}) for $n\leq -2$, since it will be divergent at 
$\tau=\frac{\mu_{0}}{\mu}$,  where the mass vanishes; note that this problem is not present for $n=0$. Therefore, we have again an apparent first order phase transition. However, one can use the $\tau$-reparametrization employed in iii), $m^2(n,\tau)= [(n-1)\gamma e^{-\gamma\tau}]^{\frac{1}{1-n}}$, with $\gamma>0$ and $n\leq 0$; this parametrization maintains the negativity of $m^2$ in the complete interval $\tau \in [0,+\infty)$, with 
$lim_{\tau\rightarrow +\infty}m^2(n,\tau)=0$.

The qualitative effect of a $\lambda$ variable is similar to the case i), hastening the deformation shown in the figure 3, or, in the case of a stationary mass, causing essentially the same deformation. The similarity is valid as well in relation to the {\it dual} versions of the theory.
\begin{figure}[H]
  \begin{center}
    \includegraphics[width=.65\textwidth]{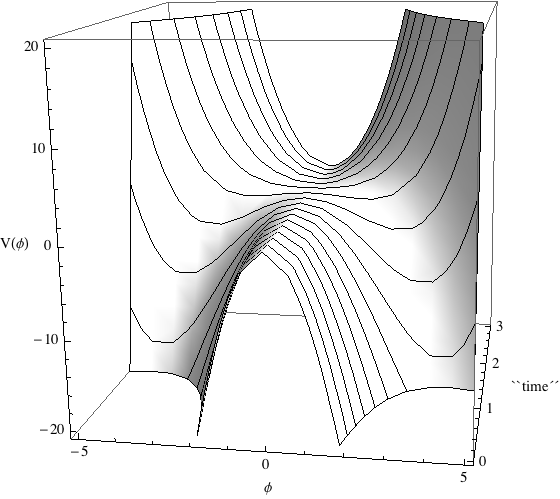}
  \caption{An apparent phase transition.}  
   \label{NOPT}
  \end{center}
\end{figure}

\noindent {\uno IV. The steepest ascent: $a=-1$}
\vspace{1em}

The analysis of these cases can be made along the lines of the previous cases; as commented, the case $a=-1$ for the steepest ascent corresponds to the backward evolution defined by the reversion of the flow parameter $\tau \rightarrow -\tau$;  in the Eq. (\ref{gradient1}) for the mass flow the reversion is equivalent to a change of sign in the constant $\mu$. Therefore, we develop the same cases i)-iv), defined by the sign of the mass at $\tau=0$; in fact,there will exist a correspondence between them. 

{\uno i)} $\mu_{0}<0$, and  $n\geq 2$, with $m^2(n, \tau =0)<0$: In this case we have a SSB scenario at $\tau=0$; negativity of $m^2$ is maintained in the interval $\tau\in [0,\frac{\mu_0}{\mu})$, and $lim_{\tau\rightarrow \frac{\mu_0}{\mu}}m^2(n,\tau)=\infty$. In the interval $(\frac{\mu_0}{\mu},+\infty)$ we have that 
$m^{2}(n,\tau)>0$, and corresponds then to an UES scenario, with $lim_{\tau\rightarrow +\infty}m^2(n,\tau)=0$; this qualitative change is shown in the figure 4.
The reparametrization used in the figure 2 works also in this case, with the gradient  $\frac{d}{d\tau}m^2= -\gamma^{2} e^{\gamma\tau}(m^2)^n$, 
which is strictly negative irrespective of the sign of $\gamma$; the solution is given by  $m^2(n,\tau)= [(n-1)\gamma e^{\gamma\tau}]^{\frac{1}{1-n}}$, which leads to a negative value for the mass at $\tau=0$ only for $\gamma<0$. Therefore, we have that   $m^2(n,\tau)<0$ in the full interval $\tau\in[0,+\infty)$ corresponding to a stable SSB scenario, y with $lim_{\tau\rightarrow +\infty}m^2(n,\tau)=-\infty$. Therefore, with this reparametrization, this case reduces to the backward evolution of the figure 1. 
\begin{figure}[H]
  \begin{center}
    \includegraphics[width=.65\textwidth]{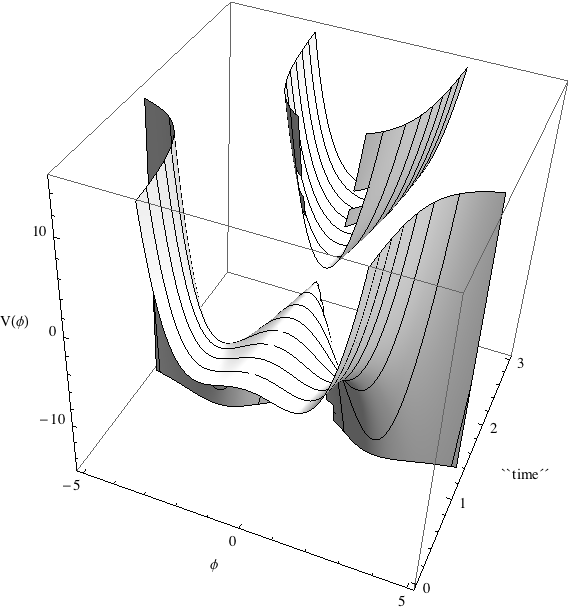}
  \caption{Note that this case corresponds to the backward evolution of the figure 2; hence, the mass gradient flow is reversible, provided that the scalar field is stationary.} 
    \label{fig:ejemplo}
  \end{center}
\end{figure}
Once the reparametrization on the mass has been made, it is easy to see the effect of a $\lambda$ variable; the condition 
$\lambda(l,\tau=0)>0$ can be preserved in the full interval $\tau\in[0,+\infty)$ with a similar reparametrization for $\lambda$, but due to the strictly negative gradient, we shall have that  $lim_{\tau\rightarrow +\infty}\lambda(l,\tau)=0$; therefore, the effect  is again to fasten the deformation on the potential caused by the (backward) mass evolution. In this case, the dual versions of the theory are in correspondence with those versions described in the section III-i).
\\

{\uno ii)} $\mu_{0}<0$, and  $n\leq 0$, with $m^2(n, \tau =0)>0$: According to Eq. (\ref{mass}), in this case we have that  $m^2(n,\tau)>0$ in the interval
$\tau\in[0, \frac{\mu_0}{\mu})$, $m^2(n,\tau=\frac{\mu_0}{\mu})=0$, and $m^2(n,\tau)<0$ in the interval $\tau\in(\frac{\mu_0}{\mu},+\infty)$; hence, there exists for all values of $n$ a qualitative change from UES scenario to SSB scenario, in this case will correspond to the backward evolution of the figure 3 of the above case iv). However,
there exists as well a problem with a vanishing mass at $\tau=\frac{\mu_0}{\mu}$ for $n\leq -2$, since the gradient (\ref{pn}) will be singular, and the backward version of the figure 3 will be valid only for $n=0$. Again, we can use the (strictly negative) reparametrized gradient for avoiding the singularities, which reads $\frac{d}{d\tau}m^2= -\gamma^{2} e^{\gamma\tau}(m^2)^n$, and restricting $\gamma<0$, the solution $m^2(n,\tau)= [(n-1)\gamma e^{\gamma\tau}]^{\frac{1}{1-n}}$ leads to $m^2(n,\tau)>0$ for the full interval $ \tau\in[0,+\infty)$, and with $lim_{\tau\rightarrow +\infty}m^2(n,\tau)=0$. This case will correspond finally to the backward version of that described in III-ii).

{\uno iii)} $\mu_{0}>0$, and  $n\geq 2$, with $m^2(n, \tau =0)>0$: hence, we have a UES scenario at $\tau=0$; due to the change of sign of $\mu$ there no exists a singularity at $\tau=\frac{\mu_0}{\mu}$, which in fact is a negative value. Positivity of $m^2$ is preserved in the full interval $\tau\in[0,+\infty)$, with  $lim_{\tau\rightarrow +\infty}m^2(n,\tau)=0$. Therefore, this case corresponds to the backward evolution of the case III-iii) above, once the reparamatrization of the later has been made.

{\uno iv)} $\mu_{0}>0$, and  $n\leq 0$, with $m^2(n, \tau =0)<0$: hence, we have a SSB scenario at $\tau=0$; there no exists a singularity in the full interval
$\tau\in[0,+\infty)$, and the negativity of $m^2$ is preserved with $lim_{\tau\rightarrow +\infty}m^2(n,\tau)=-\infty$. Therefore, this case corresponds to the backward evolution of the case III-iv), after the $\tau$-reparametrization.
The effect of a parameter $\lambda$ variable is similar to the cases discussed previously.\\

\noindent {\uno V. The parameter and fields flows as infinite series, and phase transitions}
\vspace{.5em}

In the usual scheme for the RG evolution parameters, the $\beta$-functions are perturbatively evaluated in a loop expansion, which is in principle infinite; in the present scheme based on an entropy functional, the more general gradient flows for the parameters will correspond also to infinite series, from which the expressions previously discussed can be obtained as versions truncated.

Returning to Eqs. (\ref{gradient1})-(\ref{gradient3}), we can see that those gradients determine the steepest descent and ascent for ${\cal E}$, whether a sum on $n$, and $l$ is assumed:
\begin{eqnarray}
     a \frac{d}{d\tau} m^{2} \!\! & = & \!\! \mu \sum^{\pm\infty}_{n=0} \beta_{n} \big( \frac{m^{2}}{\mu^{2}} \big)^{n}, \label{inf1} \\
     a \frac{d}{d\tau}\lambda \!\! & = & \!\! \nu \sum^{\pm\infty}_{l=0} \omega_{n} \big( \frac{\lambda}{\nu} \big)^{l}, \label{inf2} \\
     a \partial_{\tau}\phi \!\! & = & \!\! \sum_{r=1,3,5,\ldots} \sigma_{r} (\Box\phi + m^{2}\phi + \frac{\lambda}{3!} \phi^{3})^{r}, \label{inf3}
\end{eqnarray}
 where the (strictly positive) constants $\mu$, $\nu$, $\beta_{n}$, $\omega_{n}$, and $\sigma_{r}$ are introduced conveniently; if $\tau$ is in effect the energy scale, then $[\mu ]=[m]$, and $\beta_{n}$ are dimensionless constants, and similarly for the other flows. We focus again on the mass flow (\ref{inf1}), noting that the expressions will work in general for the $\lambda$-gradient. In general the flows (\ref{inf1})-(\ref{inf3}) are divergent; however, under certain restrictions we are able to find convergent solutions. Note first that the Eq. (\ref{inf1}) is separable,
 \begin{equation}
       \int \frac{d(\frac{m^{2}}{\mu^{2}})}{\sum^{\pm\infty}_{n=0} \beta_{n} (\frac{m^{2}}{\mu^{2}})^{n}} = a\frac{\tau}{\mu} + C,
      \label{separable}
\end{equation}
where $C$ is an integration constant; hence, we study now the integrability conditions of (\ref{separable}), and subsequently the possible inversion in order to find the mass as a function on $\tau$. Hence, if $M(m^{2}/\mu^{2}, a)$ is the primitive of the integral (\ref{separable}), and $M^{-1}$ the inverse, then we can express symbolically the mass as $m^{2} = M^{-1} (\tau , \mu , a, C)$, where $C$ stands for the possible integration constants; furthermore, considering a similar expression for $\lambda = \Lambda^{-1} (\tau , \nu , a,C)$, we can return to the starting point, the functional (\ref{lag}), and rewrite it as
\begin{equation}
     {\cal E} (\phi(x), m(\tau), \lambda (\tau)) = \int dx^{d} (\frac{1}{2} \partial^{i}\phi \cdot \partial_{i}\phi - \frac{1}{2} M^{-1} \phi^{2} - \frac{1}{4!} \Lambda^{-1} \phi^{4}),
     \label{lagg}
\end{equation}
this Lagrangian does not distinguish between UES and SSB scenarios, one just can choice the scenario at $`\tau=0"$; in particular, this functional can describe {\it dual} versions of the theory,  connected along the trajectories in the space of parameters; this result will be valid in the general case of a scalar field dynamical along $\tau$. 

Due to the positivity of the $\beta_{n}$'s, and the fact that $n$ is even, the possibilities of finding convergent solutions for the mass flow, and that offer possible phase transitions, are restricted; the following cases are in order.

{\uno i)}  $\beta_{n} =0$ for $n\leq -2$, and $\beta_{n}$ identified with the Euler numbers $E_{n}$ for $n\geq 0$:

Thus, the truncated sum of powers is identified with the $\sec$ function,
\begin{equation}
     \sum^{\pm\infty}_{n=0,\pm 2, \pm 4,\ldots} \beta_{n} \big( \frac{m^{2}}{\mu^{2}} \big)^{n} = \sum^{+\infty}_{n=0} \frac{E_{n}}{n!} \big( \frac{m^{2}}{\mu^{2}} \big)^{^{2n}} = \sec \big( \frac{m^{2}}{\mu^{2}} \big), \quad |\frac{m^{2}}{\mu^{2}}| < \frac{\pi}{2},
     \label{euler}
\end{equation}which, after the  integration in (\ref{separable}), and the subsequent inversion, lead to
\begin{equation}
     \frac{m^{2}}{\mu^{2}} = \arcsin (a \frac{\tau}{\mu} + C_{1}) + C_{2};
     \label{asin}
\end{equation}
with the restriction $|\frac{m^{2}}{\mu^{2}}| < \frac{\pi}{2}$ coming from (\ref{euler}); similarly we have that $| a\frac{\tau}{\mu} + C_{1}|<1$. Note first that this solution is smooth,lacks singularities; consider now the case $a=1$ with the constants $C_{1}<0$, and $C_{2}=0$, in such a way that the Eq. (\ref{asin}) looks like the figure \ref{sec}:
\vspace{1em}
\begin{figure}[H]
  \begin{center}
    \includegraphics[width=.4\textwidth]{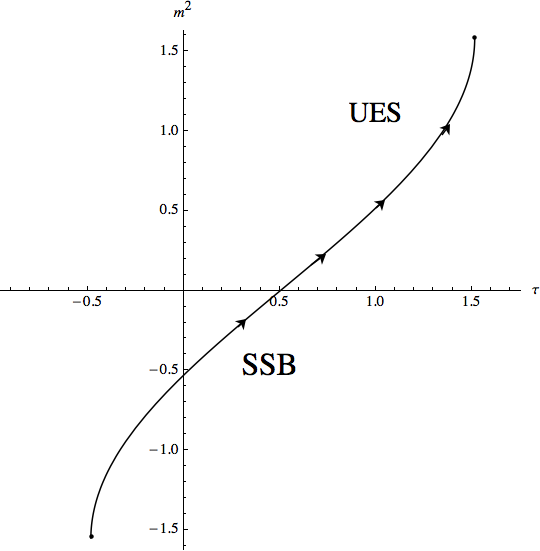}
\caption{Due to the restriction $|\frac{m^{2}}{\mu^{2}}| < \frac{\pi}{2}$, the curve only admits horizontal rigid displacement; hence the system neccesarily
transits in the form SSB-massless-UES. In this case the gradient (\ref{pn}), and its derivatives are well behaved in the interval allowed for $m^2$, and there no exist phase transitions.}    
 \label{sec}
  \end{center}
\end{figure} 
with $C_{1}<0$ and $C_{2}=0$ we have that $m^{2}(\tau =0)<0$, corresponding to a SSB scenario, which is preserved in the interval $\tau \in [0, -\mu C_{1})$; at $\tau = -\mu C_{1}$ the theory is massless, and later we have $m^{2}>0$ in the interval $\tau \in (-\mu C_{1}, -\mu C_{1} +\mu)$, corresponding to an UES scenario, with the asymptotic behavior,
\begin{equation}
\lim_{\tau\rightarrow\mu (1-C_{1})} (\frac{m^{2}}{\mu^{2}}) = \frac{\pi}{2}.
\label{massfinite}
\end{equation}
The deformation of the potential will correspond to that shown in the figure 3. Considering the property $\arcsin (-x)= -\arcsin (x)$, one can show easily that the case $a=-1$ will correspond to the curve describing a change from a UES scenario to a SSB scenario, as already described previously.
 Note that in this case, the {\it dual} versions of the theory can correspond to different scenarios; for example, a SSB scenario in the low-energy region, and an UES scenario in the high-energy region, such as described by the figure \ref{sec}; these {\it dual} versions are necessarily separated 
by a massless scenario. 

\begin{figure}[H]
  \begin{center}
    \includegraphics[width=.65\textwidth]{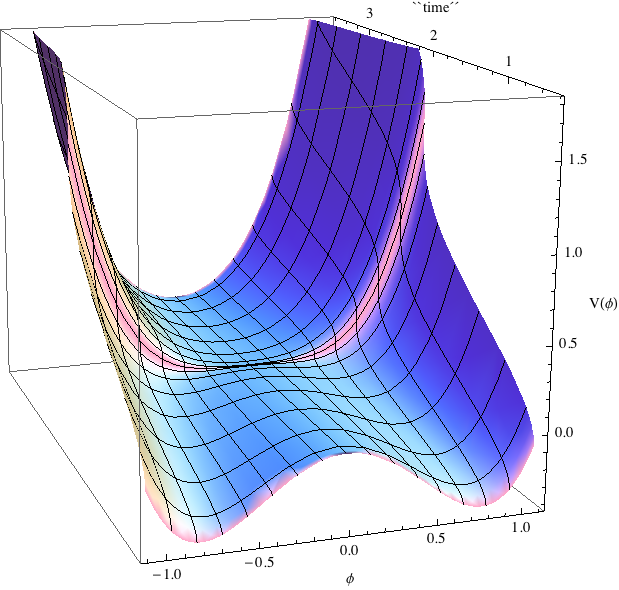}
\caption{The first order phase transition from the SSB scenario to UES scenario (a=1); this figure must be contrasted with the figure (\ref{NOPT}), that shows an apparent phase transition. In both figures we have a non-symmetry-breaking transitions, in the sense that the
reflection symmetry $\phi\rightarrow - \phi$, is preserved across the transition. Since the mass flow is reversible, the case $a=-1$ will correspond to the backward evolution, the UES domain going to the SSB domain.}    
 \label{PT1}
  \end{center}
\end{figure}

However, it is not the end of the story; considering explicitly the expansion $\frac{1}{\theta^2}\sec\theta=\frac{1}{\theta^2}(1+\frac{1}{2}\theta^2+\frac{5}{24}\theta^4+\frac{61}{720}\theta^6+...)=\frac{1}{\theta^2}+\frac{1}{2}+\frac{5}{24}\theta^2+\frac{61}{720}\theta^4+...$, that represents, in relation to the expression (\ref{separable}), a sum of the form $\sum^{+\infty}_{n=-1} \beta_{n} (\frac{m^{2}}{\mu^{2}})^{n}$, with $\theta$ identified with $\frac{m^{2}}{\mu^{2}}$,  which is admissible as a gradient flow for the mass;
this gradient flow diverges at $m^2=0$,  and represents then a first-order phase transition. Once that the integration has been made, will lead to a smooth expression for the mass, which is not invertible;
\begin{equation}
(\frac{m^{2}}{\mu^{2}}-2)\sin\frac{m^{2}}{\mu^{2}}+2\frac{m^{2}}{\mu^{2}}\cos\frac{m^{2}}{\mu^{2}}= a\frac{\tau}{\mu} + C;
\label{critical1}
\end{equation}
this expression can be considered as a correction of the expression (\ref{asin}) towards {\it criticality}; the phase transition generated by this gradient flow is illustrated in the figure (\ref{PT1}); this figure will represent qualitatively all phase transition generated by mass gradient flows. The procedure can be generalized to arbitrary order by considering the expansion $\frac{1}{\theta^{2n}}\sec\theta$, with $n=0,1,2,3..$; thus, $n$ determines  how abrupt the transition is; in particular for $n=0$ there will be not phase transition. Returning to the functional (\ref{lagg}), we have that in general $M^{-1}=M^{-1}(\tau,n)$, encoding the possibility of phase transitions, and similarly for the parameter $\Lambda^{-1}$. It is worth to note that
the {\it dual} descriptions of the theory can correspond to different phases.

The parameter $\lambda$ can take essentially the same form (\ref{asin}), and its effect is manifested, hastening, reinforcing  or generating by itself a phase transition in the case of a stationary mass; we develop in detail the case of nontrivial $\lambda$-flows in Section VI.   
   
  {\uno ii)} $\beta_{n}=0$ for $n\leq -2$ and $\beta_{n}$ for $n\geq 0$, identified with the coefficients of the expansion of $\cosh$;
the integration leads to $\arctan (\sinh \frac{m^{2}}{\mu^{2}}) = a \frac{\tau}{\mu} + C$ and hence
\begin{equation}
     \frac{m^{2}}{\mu^{2}} = \ln | \tan (a\frac{\tau}{\mu} + C_{1}) + \sec (a\frac{\tau}{\mu} + C_{1})| + C_{2}, \quad |a\frac{\tau}{\mu} + C_{1}|< \frac{\pi}{2};
     \label{cosh}
\end{equation}
the figure (\ref{cosh}) shows the behavior for $C_{2}< 0$, $C_{1}=0$, and $a=1$.

Hence, $m^{2}(\tau =0)< 0$, corresponding to a SSB scenario, which is not stable, since it will change to an UES scenario in finite $\tau$; the later will be then stable, with
\begin{equation}
     \lim_{|\tau |\rightarrow \mu\frac{\pi}{2}} \frac{m^{2}}{\mu^{2}} = \pm \infty ;
     \label{lim}
\end{equation}
therefore, the mass is not bounded, such as the previous case with  the asymptotic limit (\ref{massfinite}); due to this asymptotic behavior, there is not a choice for the constants $C_{1}$ and $C_{2}$ for which a SSB scenario is stable in the full interval (such as the previous case), it will go to an UES scenario in finite time.\begin{figure}[H]
  \begin{center}
    \includegraphics[width=.35\textwidth]{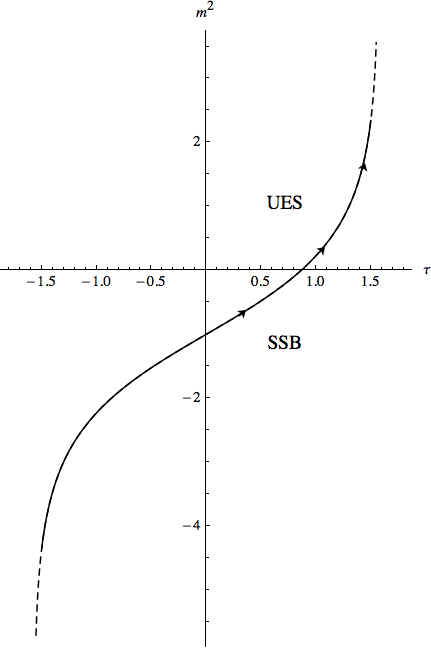}
\caption{}    
 \label{cosh}
  \end{center}
\end{figure}    
In the case with $C_{2}> 0$ (and $C_{1}=0$), we have that $m^{2}(\tau =0)>0$, corresponding to an UES scenario, which is stable in the interval $\tau >0$. Therefore, the cases with $a=1$ tend to favor UES scenarios. In the cases with $a=1$, the roles of UES and SSB scenarios will be interchanged.

 In this case involving the $\cosh$ function, the shift $\frac{1}{\theta^{2n}}\cosh\theta$ towards criticality, leads to analytical continuations for $m^2$ through poly-logarithmic functions, which do not guarantee an one-to-one correspondence  for $m^2=m^2(\tau)$, and the results are not easily interpretable. Other examples involving analytical continuations will be considered below.\\

\begin{figure}[H]
  \begin{center}
    \includegraphics[width=.5\textwidth]{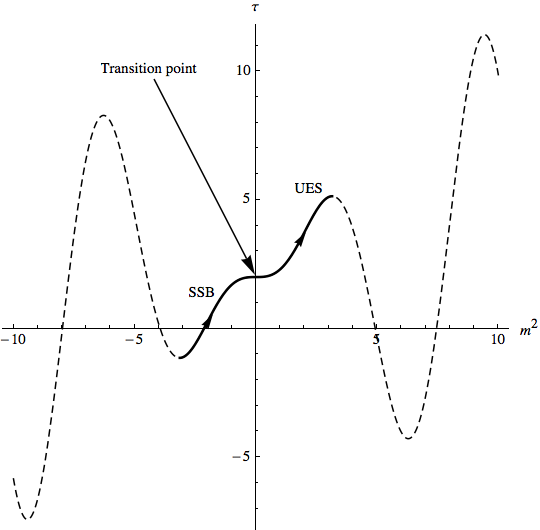}
\caption{}    
 \label{sinxcos}
  \end{center}
\end{figure}

  {\uno iii)}
Consider now first the expansion $\frac{\csc x}{x} = \frac{1}{x} (\frac{1}{x} + \frac{x}{6} + \frac{7}{360} x^{3} + \cdots )$, and fixing $\beta_{n}=0$ for $n\leq -4$, and identifying $\beta_{n}$ for $n\geq -2$ with the above expansion coefficients, we have after integration of Eq. (\ref{separable}),
\begin{equation}
     \sin (\frac{m^{2}}{\mu^{2}}) - \big(\frac{m^{2}}{\mu^{2}}\big) \cos \big(\frac{m^{2}}{\mu^{2}}\big) = a \frac{\tau}{\mu} + C, \quad |\frac{m^{2}}{\mu^{2}}| < \pi,     \label{sincos}
\end{equation}
which is not invertible; the dependence of $\tau$ on $\frac{m^{2}}{\mu^{2}}$ is shown in the figure \ref{sinxcos}, for $a=1$; the continuous line 
shows the region where the correspondence is one to one, and where is valid the restriction on $\frac{m^{2}}{\mu^{2}}$. The arrows indicate the direction of the gradient of $m^2$, and the transition from a SSB scenario to an UES scenario. 
\vspace{1em}
\vspace{1em}
\begin{figure}[H]
  \begin{center}
    \includegraphics[width=.5\textwidth]{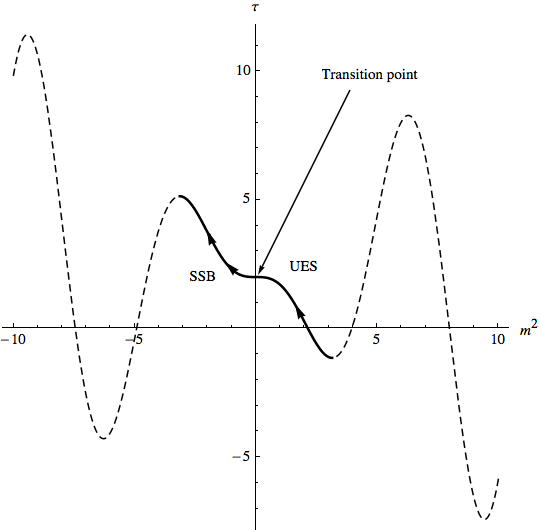}
\caption{The reversion with $a=-1$, note that the reversion does not mean to return through the curve shown
in the figure \ref{sinxcos}, since the gradient of the mass must change of sign; thus, we have a transition form an UES scenario to a SSB scenario.}    
 \label{inversefigure}
  \end{center}
\end{figure}    
In this case the gradient (\ref{inf1}) explicitly reads $a\frac{d}{d\tau}m^2=\mu\csc(\frac{m^{2}}{\mu^{2}})/(\frac{m^{2}}{\mu^{2}})$, which diverges at $m^2=0$ corresponding to the massless scenario, and induces a divergence in the gradient (\ref{var-2}); hence, we have a {\it first order} phase transition. In fact, all-order derivatives diverge at the transition point; the generalized expansion is given in this case by $\frac{1}{\theta^{2n+1}}\csc\theta$, with $n= 0,1,2,..$.  The figure (\ref{PT1}) describes the phase transition at the level of the potential.

{\uno iv)} Analytical continuations for $m^2(\tau)$
\vspace{.4em}

In all previous cases we describe stable UES and SSB scenarios after the transition, and  
the system can not return to the original scenario; as we shall see in this section, when criticality is absent, such a returning will be possible only for analytical continuations for $m^2$.

\begin{figure}[H]
  \begin{center}
    \includegraphics[width=.45\textwidth]{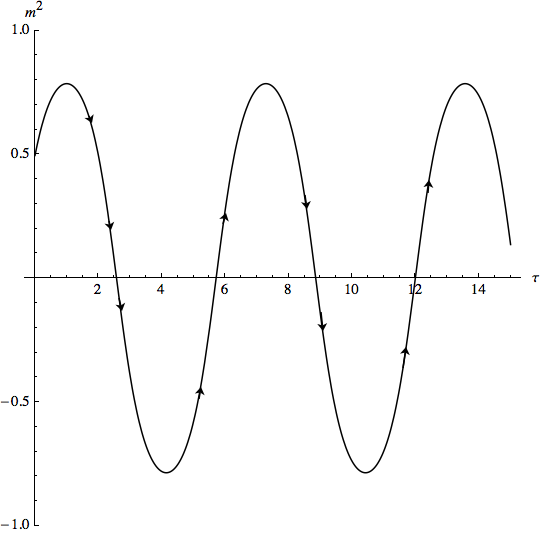}
\caption{The real and imaginary parts of $m^2$ ; with the appropriate values of the constants, a {\it raising} of the curve will lead to a stable UES scenario with $Re(m^{2})>0$, or $Im(m^{2})>0$ in the full interval, and similarly for the SSB scenario . In this figure the constants have been chosen in such a way that periodic UES-SSB transitions are present; the figure \ref{im-periodic} shows these transitions at the level of the evolution of the potential $V(\phi )$.}    
 \label{oscilation}
  \end{center}
\end{figure}

Consider the complex function $\cos z = 1 - \frac{z^{2}}{z!} + \frac{z^{4}}{4!} - \frac{z^{6}}{6!} +\cdots$, with $z= x+\imath y$, and $|z|< \infty $; with the restriction $x=0$, the function reduces to $\cos \imath y= 1+ \frac{y^{2}}{2!} + \frac{y^{4}}{4!} + \frac{y^{6}}{6!} +\cdots$. where all expansion coefficients are strictly positive; hence with the restriction $\beta_{n}=0$ for $n\leq 2$, we have $ \sum^{+\infty}_{n\geq 0} \beta_{n} \big( \frac{m^{2}}{\mu^{2}} \big)^{n} = \cos \imath \frac{m^{2}}{\mu^{2}}$, where we have identified $\beta_{n}$ for $n\geq 0$, with the above expansion coefficients.
Therefore, the integration (\ref{separable}) leads to a complex expression for $m^{2}$, where $C_{1}$ and $C_{2}$ are constants,
\begin{equation}
     \frac{m^{2}}{\mu^{2}} = -2\imath \arctan [ C_{2} e^{\imath (a\frac{\tau}{\mu}+C_{1})}] + \imath \frac{\pi}{2};
     \label{complexmass}
\end{equation} 
now, restricting $C_{2} \in R$, one can find the decomposition in parts real and imaginary,
\begin{equation}
     \frac{m^{2}}{\mu^{2}} = \frac{1}{2} \ln \frac{C^{2}_{2} +2 C_{2} \sin (a\frac{\tau}{\mu}+ C_{1}) +1}{C^{2}_{2} - 2C_{2} \sin (a\frac{\tau}{\mu} +C_{1})+1} + \imath \big\{\pi (\frac{1}{2} -2k) - \arctan \frac{2C_{2} \cos (a\frac{\tau}{\mu} +C_{1})}{1-C^{2}_{2}}\big\},
     \label{complexmasss}
\end{equation}
where $k$ is an integer, and $C^{2}_{2} \neq 1$.  In spite of their appearance, both parts have essentially the same qualitative behavior,which is shown in the figure \ref{oscilation}; the case of $\lambda$ variable is similar to those described previously. Note that with this analytical continuation, the role of  the parameter $a$ is secondary, without a qualitative effect in the behavior of $m^2$.
\begin{figure}[H]
  \begin{center}
    \includegraphics[width=.5\textwidth]{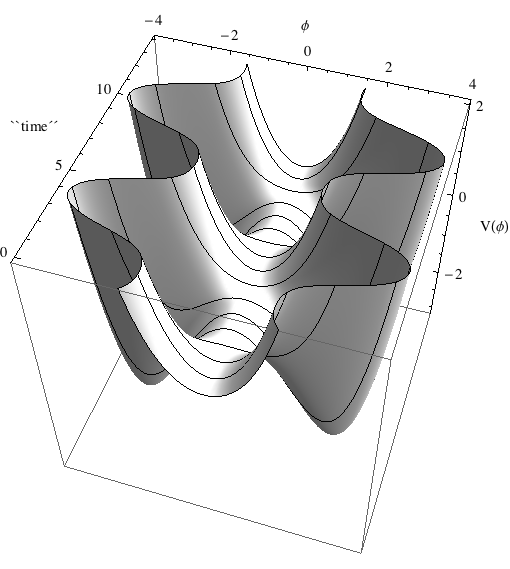}
\caption{The periodic deformations of the potential $V(\phi)$ for both real and imaginary parts of $m^2$; if the potential starts at $\tau=0$ in an UES scenario, then it returns to a such scenario in finite time, but crossing by a SSB scenario. These transitions are not phase transitions such as that shown in the figure \ref{PT1}; however, represent similarly  non-symmetry-breaking deformations, since the symmetry $\phi\rightarrow -\phi$ is preserved intact across the transition.}    
 \label{im-periodic}
  \end{center}
\end{figure}   
Generalizing the expansion to $\frac{1}{y^n}\cos(iy)$, with $n=0,2,4,...$, {\it criticality} emerges from $n\geq 2$, with the massless scenario as the UES--SSB phase transition point; when {\it criticality} appears, the periodic behavior shown in the figure \ref{im-periodic} 
disappears abruptly; the figure \ref{PT1} is representative of such a transition.  The solutions for $m^2=m^2(\tau,n)$, are given by the Poly-logarithmic functions of order $n+1$.
There exist other complex functions that allow different analytical continuations for $m^2$, but they will be considered in futures works.

{\uno v)} Other cases.

The expansion $\frac{\tan x}{x} = \frac{1}{x} (x + \frac{x^{3}}{3} + \frac{2}{15} x^{5} + \cdots )$, allows describe a case similar to V.ii), with $\beta_{n}=0$ for $n\leq 2$ and $\beta_{n}$ for $n\geq 0$ identified with the above expansion coefficients; therefore, $\int\frac{m^{2}}{\mu^{2}} \cdot \cot \frac{m^{2}}{\mu^{2}} d(\frac{m^{2}}{\mu^{2}}) = \frac{m^{2}}{\mu^{2}}\ln(1-\exp{2 i  \frac{m^{2}}{\mu^{2}}})-\frac{i}{2} [( \frac{m^{2}}{\mu^{2}})^2+Li_{2}(\exp{2 i  \frac{m^{2}}{\mu^{2}}})]  = a \frac{\tau}{\mu} + C$,  where $Li_{2}$ is a Poly-logarithmic function, and $|\frac{m^{2}}{\mu^{2}}| <\pi/2$;  as opposed to the previous case, there is no a phase transition of any order, since all-order derivatives of the mass (and consequently of the functional ${\cal E}$) are smooth in the domain of the expansion. Generalizing the expansion to $\frac{1}{\theta^{2n+1}}\tan\theta$,  with $n=0,1,2,..$, {\it criticality} emerges from $n\geq1$, and the critical point is again
the massless scenario, separating UES and SSB phases.

Considering now the expansion $\theta^{2n+1}arccoth\theta = \theta^{2n+1} (\frac{1}{\theta} + \frac{1}{3\theta^{3}} + \frac{1}{5\theta^{5}} + \cdots )$, with $n=0,\pm 1, \pm2, \pm 3,..$, and $|\theta| >1 $, allows describe the case with $\beta_{n} =0$ for $n\geq 0$ and $\beta_{n}$ for $n\leq -2$ identified with the expansion coefficients. This expansion leads to the non-integrable expression
\[
     \int \frac{(\frac{m^{2}}{\mu^{2}})^{-2n-1} d(\frac{m^{2}}{\mu^{2}})}{arcctanh (\frac{m^{2}}{\mu^{2}})} = 2 \int \frac{(\frac{m^{2}}{\mu^{2}})^{-2n-1}  d(\frac{m^{2}}{\mu^{2}})}{\ln [\frac{\frac{m^{2}}{\mu^{2}}+1}{1-\frac{m^{2}}{\mu^{2}}}]} = a \frac{\tau}{\mu} +C;  \quad |\frac{m^{2}}{\mu^{2}}| >1,\nonumber
\] 
due to the restrictions on the values of $m^2$,  the value $m^2=0$ is not allowed, and a transition between SSB-UES scenarios is not possible; hence, an UES scenario at $\tau=0$, is stable for all interval, and similarly for the SSB case; there is no {\it criticality} for any value of $n$.\\

\noindent {\uno VI. The $\lambda$-gradient flow and double phase transitions}
\vspace{.5em}

A $\lambda$-gradient flow can connect along trajectories in the parameters space the dual versions of the theory; additionally  can hasten and/or reinforce a phase transition induces by a mass flow, provided that both flows are in {\it phase}. However, this is not always the case; we consider in this section the diverse situations with a non-stationary self-interacting parameter $\lambda$, first {\it in phase} with the mass flow, and at the end  {\it out of phase}.

{\uno i)} gradient flows {\it in phase}.

\begin{figure}[H]
  \begin{center}
    \includegraphics[width=.55\textwidth]{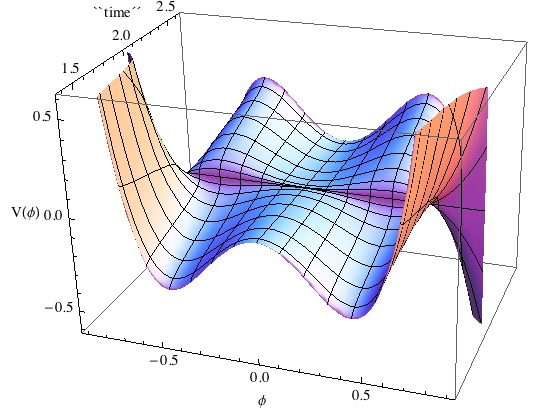}
\caption{ A first order phase transition with flows in phase; the transition consist of ``turning upside" the original potential related with the SSB scenario. As opposed to the original phase, the energy in the resultant phase is not bounded from below; another effect of the transition is the transit of vacuum from two lowest-energy states to one lowest-energy state. In relation to the sign of the parameters, the transition is such that $(m^2<0,\lambda>0)\rightarrow(m^2=0,\lambda=0)\rightarrow(m^2>0,\lambda<0)$. Since the flows are reversible, the case with $a=-1$ corresponds to the inverse description.}    
 \label{phase1}
  \end{center}
\end{figure}

Consider by concreteness the solution (\ref{critical1}) for $a=1$, which works as solution for gradient flows for the mass and the parameter $\lambda$; furthermore, we assume that the constant $C$ on the right hand side is the same for both flows, thus the evolution for the pair $(m^2,\lambda)$ is {\it in phase}; hence, one can tune the flows in order to have a SSB scenario at $\tau=0$, as illustrated in the figure \ref{phase1}. Since the evolutions is in phase, the pair  $(m^2,\lambda)$ reaches the critical point $(0,0)$ to the 
{\it same time}, reinforcing the phase transition. Note that the resultant phase is not that shown in the figure \ref{PT1}  with a parameter  $\lambda$ fixed. In the transition point $(m^2=0,\lambda=0)$,  the potential vanishes $V=0$, and is represented by a horizontal line in the figure; in such a point, the theory is massless and free, and the Lagrangian is reduced to the kinetic part. If one choices the criticality parameter $n=0$, in such a way that there is no a phase transition, then all physical quantities are smooth at the point $(m^2=0,\lambda=0)$.

In this manner, we return again to the functional (\ref{lagg}), which has to this point the possibility of phase transitions with flows {\it in phase}. In this example we have considered the same solution  (\ref{critical1}) for both flows; however, one can choice for the $\lambda$-flow the solution (\ref{sincos}), for instance, and then to tune the flows; these other possibilities will lead to qualitatively different transitions that those shown in the figure \ref{phase1}.  In the next sub-section we shall consider an example of a phase transition with flows {\it out of phase}.
{\uno ii)} gradient flows {\it out of phase}.

\begin{figure}[H]
  \begin{center}
    \includegraphics[width=.55\textwidth]{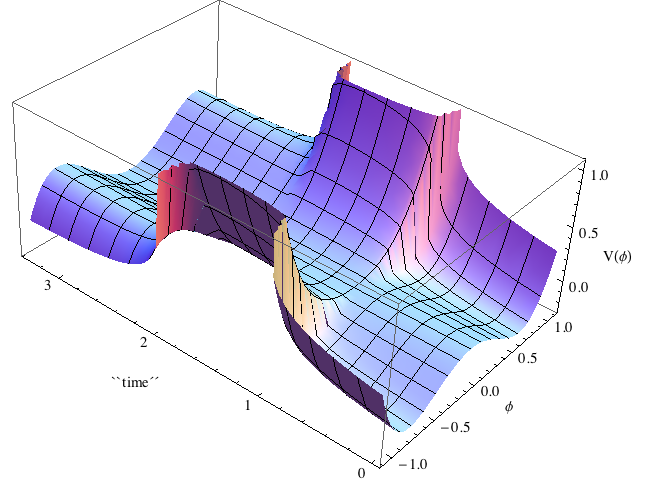}
\caption{  First order phase transitions with flows out of phase. In the original, and intermediate phases, the energy 
is bounded from below, as opposed to the resultant phase; note that the zero-energy point is an unstable maximum
in the first phase, and a stable minimum in the intermediate and final phases. The vacuum transits  from two lowest-energy states in the first phase, to one lowest-energy state in the last two phases. In relation to the sign of the parameters, the transition is such that $(m^2<0,\lambda>0)\rightarrow(m^2=0,\lambda>0)\rightarrow(m^2>0,\lambda>0)\rightarrow(m^2>0,\lambda=0)\rightarrow(m^2>0,\lambda<0)$, where the transition points correspond to the massless and free scenarios.}    
 \label{outphase1}
  \end{center}
\end{figure}

\begin{figure}[H]
  \begin{center}
    \includegraphics[width=.55\textwidth]{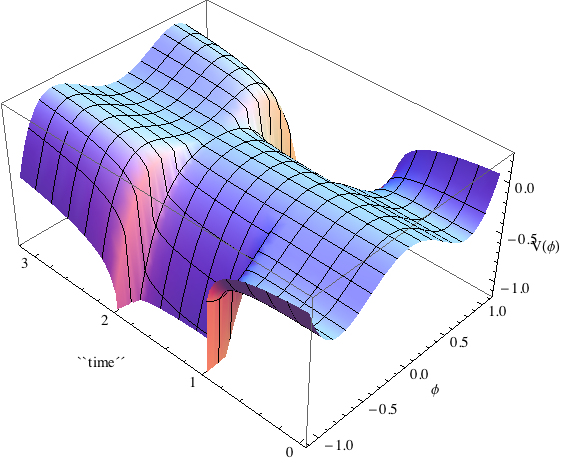}
\caption{ Starting in a SSB phase at $\tau=0$, we have a first phase transition generated by the $\lambda$-flow, and then a second phase transition generated by the mass flow, {\it i.e.}, in the inverse order described in the figure \ref{outphase1}.}    
 \label{outphase2}
  \end{center}
\end{figure}

Along the same lines we can consider the solution (\ref{critical1}) for $a=1$, but choosing different values for the constant $C$; hence the flows will be {\it out of phase}. In the figure \ref{outphase1} the choosing is such that the mass flow generates first a phase transition, and a moment later, the $\lambda$-flow generates the second phase transition. In the figure \ref{outphase2}, the roles are interchanged, in such a way that the $\lambda$-flow generates the first phase transition. In order to compare with the case of flows {\it in phase} illustrated in the figure \ref{phase1} , we have chosen 
a SSB scenario at $\tau=0$; note that in the three figures the resultant phase is the same. 

Once the two phase transitions have occurred, the final phase is stable as $\tau\rightarrow + \infty$, and there no exist anymore phase transitions; in general there will exist as many phase transitions as parameters defining the theory. To this point, we can incorporate phase transitions with
gradient flows {\it out of phase} to the functional (\ref{lagg}); in the next section we develop the case of evolving $\phi$-field, the unique ingredient maintained stationary along the $\tau$-gradient to this point, and that finally we shall incorporate to the functional (\ref{lagg}).


\noindent {\uno VII.The reaction-diffusion for the field $\phi$}

To this point we have taken advantage on the fact that the gradient flows (\ref{gradient1}), and (\ref{gradient2}) for $m^{2}$ and $\lambda$ are decoupled, and are independent on the field $\phi (x,\tau )$, which was maintained stationary. We turn now towards the gradient (\ref{gradient3}), for the field $\phi$, which includes explicitly $m^{2}(\tau )$, and $\lambda(\tau )$. As a {\it heat}-like non-linear equation, $a$ will correspond to the diffusivity constant, and the mass and self-interacting term will correspond to {\it reaction} terms. Non-linear reaction-diffusion equations play a key role in a great number of models of heat and reaction-diffusion process, from biology, chemistry, genetics and many others. We use the literature available on these differential equations, in particular the reference \cite{poly}. In this paper we develop only the one space dimension case;  although one may think that is only of pedagogical interest, will show interesting properties in the present context.
Additionally, some aspects of the one-dimensional case, will can generalize to any dimension in a direct way.
 We consider the case with $a=1$, and with $m^{2}$ and $\lambda$ fixed; explicitly the equation reads
\begin{equation}
     \partial_{\tau} \phi(x,\tau) = \big( \frac{\partial}{\partial x^{2}} + m^{2}\big) \phi(x,\tau ) + \frac{\lambda}{3!} \phi^{3} (x,\tau );
     \label{one-dim}
\end{equation}
this equation admits different solutions,and we consider first the traveling-waves solutions.\\

\noindent {\uno VII.I Traveling-waves solutions}
\begin{equation}
     \phi (x,\tau) = [ b C_{vac} + C exp^{(-3 \frac{m^{2}}{2}\tau +b \sqrt{\frac{m^{2}}{2}} x)} ]^{-1},
     \label{traveling}
\end{equation}
where $C_{vac} =  \sqrt{-\frac{\lambda} {6m^{2}} }$, is essentially the inverse of $\phi_{vac}$, the position of the two lowest-energy states in the SSB scenarios; $C$ is a completely arbitrary constant, and $b=\pm 1$, with $b=1$ for the {\it advanced} waves, and $b=-1$ for the {\it retarded} ones. The properties of the solution, depend sensitively on the sign of $m^2$; two cases are in order.\\

\noindent {\uno VII.Ia) SSB case; $m^2<0$} 

If we choice $m^{2}<0$, for particularizing the analysis to SSB scenarios, then the factor for the spatial coordinate necessarily is pure imaginary, essentially the usual taquionic mass induce by SSB; additionally the constant $C_{vac}$ is real; therefore, the existence of reaction-diffusion traveling waves for the $\phi$-field implies necessarily its analytical continuation through harmonic functions in the $x$-
coordinate. Hence, we write the field $\phi$ in terms of real and imaginary parts $\phi = \varphi (x,\tau ) + \imath \eta (x,\tau )$, which can be determined from Eq. (\ref{traveling}); the explicit dependence of $\varphi$ and $\eta$ in the coordinates $(x,\tau )$ is shown in the figures (\ref{rtws}) and (\ref{itws}) below.
For values of $\tau$ far from the singularities, the diffusive evolution of the fields is as expected, the peaks hight decay for long $\tau$.

In this manner, the original Lagrangian (\ref{lag}) becomes an analytic functional on the complex field $\phi$ (it does not depend on $\overline{\phi}$); hence, since an analytic functional can not attain relative minimums and/or maximums (although the saddle points are possible), it seems to be that the SSB scenarios have been lost. However, the 
 two minimum-energy states scenarios are present always in the $\varphi$-direction, but they are not  stable relative minimums. We can realize that as follows; the complex Lagrangian (\ref{lag}) must be decomposed into real and imaginary parts, and later one must consider each part in the real domain. The Lagrangian decomposed reads
\begin{eqnarray}
     {\cal E} \!\! & = & \!\! \int \frac{1}{2} (\partial_{x}\varphi )^{2} - \frac{1}{2} (\partial_{x}\eta )^{2} -V(\varphi , \eta ) \label{reallag} \\
     \!\! & + & \!\! \imath \int \eta [\frac{1}{2} \Box\varphi + \frac{1}{2} m^{2}\varphi + \frac{\lambda}{3!} \varphi^{3}] + \varphi [\frac{1}{2} \Box\eta + \frac{1}{2} m^{2}\eta - \frac{\lambda}{3!} \eta^{3}], \label{imaglag} \\
     V(\varphi , \eta ) \!\! & = & \!\! \frac{1}{2} m^{2} (\varphi^{2} -\eta^{2}) + \frac{\lambda}{4!} (\varphi^{4} + \eta^{4} -6\varphi^{2} \eta^{2}); \label{potreallag}
\end{eqnarray}
in the real part (\ref{reallag}), $V(\varphi , \eta )$ includes the mass terms for the fields (both have the same mass), the $\lambda\varphi^{4}$ and $\lambda\eta^{4}$ self-interacting terms, and the coupling term of the form $\varphi^{2}\eta^{2}$. Furthermore, from the explicit expressions for the fields $(\varphi,\eta)$, we have that $\varphi(\tau,-x)=\varphi(\tau,x)$, and $\eta(\tau,-x)=-\eta(\tau,x)$; hence  the imaginary part (\ref{imaglag}) is odd under $x\rightarrow-x$, and vanishes identically, since the integration is over all space, $\int^{+\infty}_{-\infty}d x$.
Hence, the real Lagrangian describes the dynamics of two scalar fields interacting to each other, with the symmetry $\varphi\rightarrow -\varphi$, and independently $\eta\rightarrow -\eta$. Furthermore, under the condition $\eta =0$,  the real part (\ref{reallag}) will reduce to the original Lagrangian (\ref{lag}) for a real scalar field.

\vspace{.5em}
\begin{figure}[H]
  \begin{center}
    \includegraphics[width=.55\textwidth]{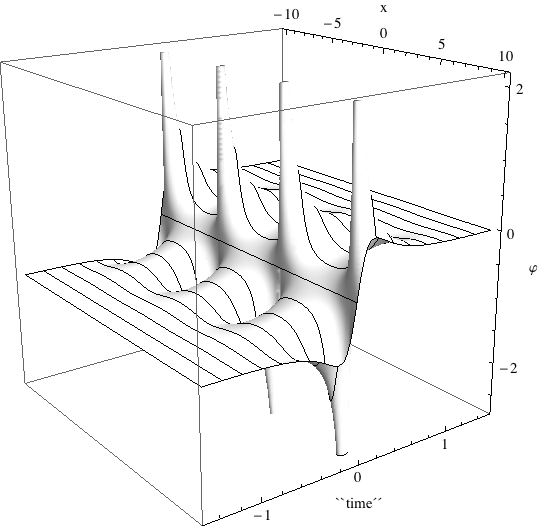}
\caption{Real part of the solution (\ref{traveling}), $\varphi=[C exp^{-3 \frac{m^{2}}{2}\tau} \cos( \sqrt{\frac{|m^{2}|}{2}} x)+b C_{vac}]/ [C^2exp^{-3m^2\tau}+2b CC_{vac} exp^{-3 \frac{m^{2}}{2}\tau} \cos( \sqrt{\frac{|m^{2}|}{2}} x )+C_{vac}^2] $; this expression diverges at two values of the $\tau$-parameter, according to the denominator, and due to the presence of harmonic functions, the singularities are periodic in the spatial direction. The asymptotic limits are $lim_{\tau\rightarrow-\infty}\varphi(x,\tau)=b/C_{vac}$ (essentially $\phi_{vac}$), and $lim_{\tau\rightarrow+\infty}\varphi(x,\tau)=0$. The singularities appear for $\lambda \neq 0$, {\it i. e.}, as a {\it reaction} effect.}
 \label{rtws}
  \end{center}
\end{figure}  
\vspace{.5em}
\begin{figure}[H]
  \begin{center}
    \includegraphics[width=.55\textwidth]{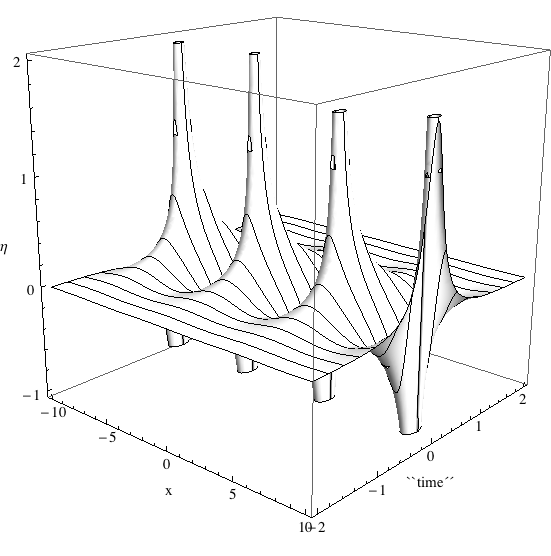}
\caption{Imaginary part of the solution (\ref{traveling}), $\eta = C exp^{-3 \frac{m^{2}}{2}\tau} \sin(b \sqrt{\frac{|m^{2}|}{2}} x)/ [C^2exp^{-3m^2\tau}+2b CC_{vac} exp^{-3 \frac{m^{2}}{2}\tau} \cos( \sqrt{\frac{|m^{2}|}{2}} x )+C_{vac}^2] $; in this case the field diverges  only  for a  value of  the $\tau$-parameter; the asymptotic limits are $lim_{\tau\rightarrow\pm\infty}\eta(x,\tau)=0$.}
 \label{itws}
  \end{center}
\end{figure} 
The singularities shown in the figures (\ref{rtws}) and (\ref{itws}) at the level of the fields $(\varphi,\eta)$, will imply singularities in the functional (\ref{reallag}), and all its $\tau$-derivatives, representing then obvious phase transitions for the theory; explicitly we have that $\partial_{\tau}{\cal E}=\int^{+\infty}_{-\infty}d x[(\partial_{\tau}\eta)^2-(\partial_{\tau}\varphi)^2]$. 

\begin{figure}[H]
  \begin{center}
    \includegraphics[width=.55\textwidth]{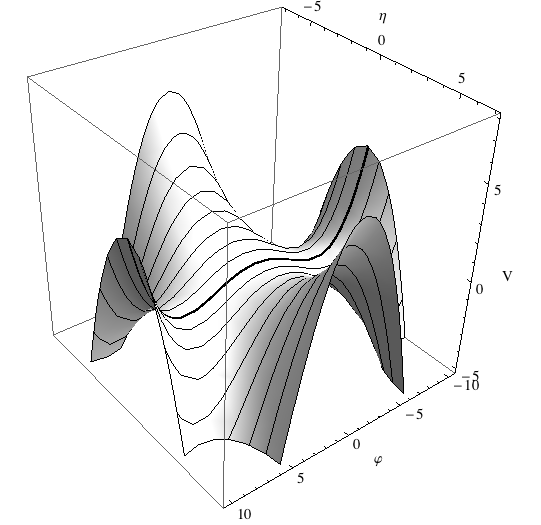}
\caption{
configuration of $V(\varphi,\eta)$, at a fixed $\tau$, far from singularities described in the figures \ref{rtws}, and \ref{itws}; in bold-face, the configuration of $V(\varphi,\eta=0)$, which can be considered, for $\varphi$-direction displacements,  as the ``usual" configuration for the potential with an unstable maximum point, and two ``stable" minimum points; however, in the present description, those three critical points correspond to saddle points. Therefore, there exists no a stable vacuum to choice for expanding around it.}
 \label{RPPOTENTIAL}
  \end{center}
\end{figure}

We analyze now the maximums, minimums, and saddle points for the potential $V$. The critical points (CP) are $CP(V)= \{ (\varphi =0, \eta =0); (\varphi = \pm\sqrt{3!(\frac{-m^{2}}{\lambda})}, \eta =0)\}$, and its Hessian matrix
\begin{equation}
    {\cal H} (V) = \left( \begin{array}{cc}
    m^{2} + \frac{\lambda}{2} (\varphi^{2}-\eta^{2}) & -\lambda\varphi\eta  \\
    -\lambda\varphi\eta & -m^{2}+ \frac{\lambda}{2} (\eta^{2} - \varphi^{2}) 
    \end{array} \right);
    \label{hessian}
\end{equation}
hence, $\det {\cal H} (\varphi =0, \eta =0) =-m^{4}<0$, and we have a saddle point. Similarly, $\det {\cal H} (\varphi =\pm\sqrt{3!(\frac{-m^{2}}{\lambda})}, \eta =0) =-8m^{4}<0$, and we have again two saddle points; the $V$-potential is shown in the figure (\ref{RPPOTENTIAL}). In general, the determinant of ${\cal H}$ is a strictly negative functional of $\tau$, $det ({\cal H})=-[m^2+\frac{\lambda}{2}(\varphi^2-\eta^2)]^2-\lambda^2\varphi^2\eta^2$, and the trace an invariant under the $\tau$-flow, $Tr({\cal H})=0$.

As a conclusion of this section, we can say that the deformation of the theory through `traveling waves" generates a {\it doubling} of the fields, and the deformed version can be reduced to a fully real scheme; however, the appearance of the second field $\eta$ induces an instability in the degenerate vacuum, since there is no exists a stable minimum around which to expand the theory. The doubling of the fields will lead inevitably to a doubling as well in the field variables in the phase space, and and the Hilbert space of states; however, the effects of the deformations of the theory on its quantum formulation will be considered in future works.

\noindent {\uno VII.Ib) UES case; $m^2>0$} 

In this case the analytical continuation for the field $\phi$ appears by mean of the constant $C_{vac} =  \sqrt{-\frac{\lambda} {6m^{2}} }$, which is pure imaginary; the factor for the spatial coordinate is now real, and hence there is a drastic change respect to the SSB case discussed above. For simplicity we consider that the arbitrary constant $C$  is real;  the real $(\varphi)$, and imaginary part $(\eta)$ for the field $\phi$ read,
\begin{eqnarray}
     \varphi (x,\tau ) \!\! & = & \!\!C\frac{exp(-\frac{3}{2} m^{2}\tau+b\sqrt{\frac{m^2}{2}}x)}{|C_{vac}|^2+ C^2exp(-3 m^{2}\tau+2b\sqrt{\frac{m^2}{2}}x )}, \label{ues-solitary} \\
     \eta(x,\tau )\!\! & = & \!\! b\frac{|C_{vac}|}{|C_{vac}|^2+ C^2exp(-3 m^{2}\tau+2b\sqrt{\frac{m^2}{2}}x )}, \label{ues-solitary2} 
\end{eqnarray}
with the asymptotic $lim_{\tau\rightarrow\pm\infty}\varphi(\tau,x)=0$,  $lim_{\tau\rightarrow -\infty}\eta(\tau,x)=0$, and  $lim_{\tau\rightarrow +\infty}\eta(\tau,x)=\frac{b}{C_{vac}}$; as opposed to the previous case related to SSB scenarios, these field configurations do not show singularities; hence, there no exist phase transitions. Furthermore, another important difference is that in this case the Lagrangian will have a non-vanishing imaginary part, since in this case the odd-character under $x\rightarrow-x$, is not present. Considering
 the imaginary part (\ref{imaglag}), the terms of the form $\eta\Box\varphi$, and $\varphi\Box\eta$ can be considered as the kinetic terms, and thus we shall have a potential that includes self-interacting and coupling terms, $U(\varphi
, \eta) \equiv \eta (\frac{1}{2} m^{2}\varphi + \frac{\lambda}{3!}\varphi^{3}) + \varphi (\frac{1}{2} m^{2}\eta - \frac{\lambda}{3!} \eta^{3})$.

Hence, both parts of the Lagrangian describe the dynamics of two scalar fields interacting to each other, in principle in an UES scenario; we shall see, however, that  the original UES scenario has been lost in some sense. The non-vanishing imaginary part (\ref{imaglag}) has the symmetry $(\varphi ,\eta)\rightarrow (-\varphi , -\eta)$. Furthermore, under the condition $\eta =0$, the imaginary part (\ref{imaglag}) vanishes, and the real part (\ref{reallag}) will reduce to the original Lagrangian (\ref{lag}) for a real scalar field.

\begin{figure}[H]
  \begin{center}
    \includegraphics[width=.55\textwidth]{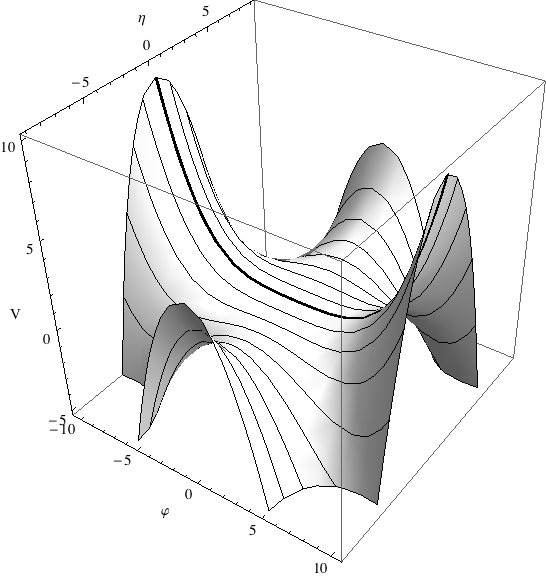}
\caption{configuration of $V(\varphi,\eta)$, at a fixed $\tau$ with $m^2>0$; in bold-face, the configuration of $V(\varphi,\eta=0)$, which can be considered, for $\varphi$-direction displacements,  as the ``usual" configuration for the potential with a stable minimum point; however, in the present description, that critical point corresponds to a saddle point.}
 \label{UESPOT}
  \end{center}
\end{figure}

We analyze now the maximums, minimums, and saddle points for the potential $V$ with $m^2>0$; the CP are now, $CP(V)= \{ (\varphi =0, \eta =0); (\varphi=0,\eta = \pm\sqrt{3!(\frac{m^{2}}{\lambda})})\}$, where the roles of the fields $\varphi$, and $\eta$ turn out to be interchanged respect to the case with $m^2<0$ discussed above;  its Hessian matrix is given by the expression (\ref{hessian}), and leads again to three saddle points; the potential is illustrated in the figure \ref{UESPOT}.
Physically the potential shown in the figure \ref{UESPOT}, is equivalent to that shown in figure \ref{RPPOTENTIAL}; the potentials can be related by a simple interchange of the fields, $\varphi\leftrightarrow\eta$; additionally note that the potential $V$ is invariant under the transformation $(\varphi\leftrightarrow\eta, m^2\rightarrow-m^2)$. Therefore, at the level of the potential $V$, the scenarios SSB and UES are not distinguishable. The difference is in the presence of the Imaginary part  for the Lagrangian for UES scenarios; in particular with an additional potential $U$.

\begin{figure}[H]
  \begin{center}
    \includegraphics[width=.55\textwidth]{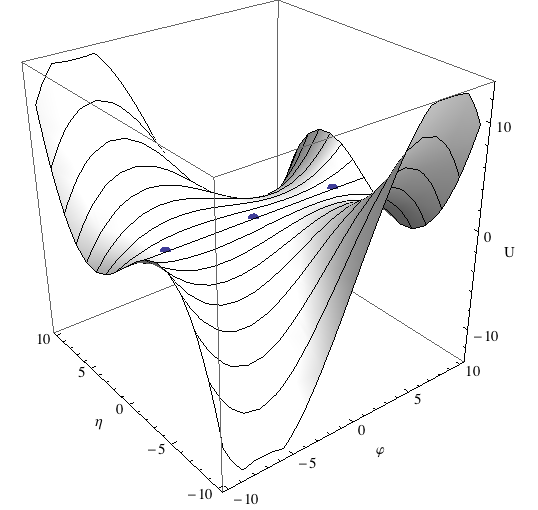}
\caption{ configuration of $U(\varphi,\eta)$, at a fixed $\tau$; the three saddle points have $\eta=0$, and 
$U(\varphi, \eta=0)$=0, condition that represents in fact the complete line containing those points.}
 \label{IMPPOTENTIAL}
  \end{center}
\end{figure}

One can show that the $U$-potential has a Hessian matrix with entries given by  ${\cal H}_{11}(U)=-{\cal H}_{12}(V)=-{\cal H}_{22}(U), {\cal H}_{12}(U)={\cal H}_{11}(V)={\cal H}_{21}(U)$; therefore, both matrices have the same determinant, $det{\cal H}(U)=det{\cal H}(V)$, and $Tr {\cal H}(U)=0$.
In spite of that the potentials $U$, and $V$ look as different, both have the same critical points $CP(U)=CP(V)$, and thus one can show that $CP(U)$ correspond also to saddle points; the potential is shown in the figure (\ref{IMPPOTENTIAL}). 

Usually the appearance of imaginary contributions to the  potential can be associated directly to instabilities; however, in this case, the same real contribution develops instabilities of vacuum.
In the next section we develop the case of self-similar solutions, in which the SSB scenario is maintained under $\tau$-evolution; hence, a stable degenerate vacuum will be present along the deformation of the theory.

\noindent {\uno VII.II Solitons; self-similar solutions} 

\begin{figure}[H]
  \begin{center}
    \includegraphics[width=.6\textwidth]{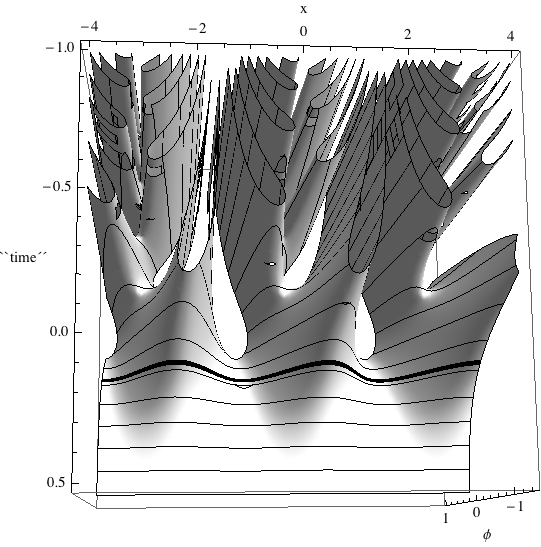}
\caption{In bold-face, the curve that represents the field configurations at $\tau=0$; such configurations can be identified with the fixed points related with the solution space of the classical equations of motions. For $\tau>0$, the diffusive evolution with the expected effect on the field, which is smoothed, with the  asymptotic $lim_{\tau\rightarrow+\infty}\phi(x,\tau)=0$; the field is periodic in the spatial direction, due to the presence of harmonic functions. For $\tau<0$, the dramatic effect of the irreversibility in the diffusive evolution; the self-similarity becomes manifest in this domain;  additionally the field configurations show a periodic pattern in the $x$-direction for certain $x$-intervals.}
 \label{ffractal}
  \end{center}
\end{figure}

\begin{figure}[H]
  \begin{center}
    \includegraphics[width=.6\textwidth]{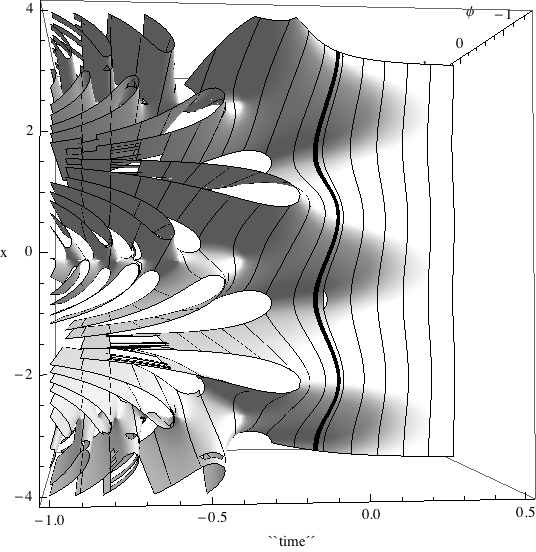}
\caption{This figure represents a slightly different angle of view, respect to the figure \ref{ffractal}; the self-similarity in the domain $\tau<0$ is manifest with the appearance of  {\it branching self-similar patterns}, which are replicated at infinitum, as $\tau\rightarrow -\infty$. The branching phenomenon is {\it left-handed} for $b=1$, since the left-branch is that undergoing the bifurcation into two branches; the right-brach generated in each pattern, continues to infinity without bifurcation; in particular, the first branch generated in the backward evolution is a right-one, and corresponds to an {\it enveloping} branch for the self-similarity pattern formation for moments later. Similarly the case with $b=-1$, is {\it right-handed}; thus, the value of $b$ determines the {\it quirality} of the backward evolution.}
 \label{fffractal}
  \end{center}
\end{figure}

\begin{figure}[H]
  \begin{center}
    \includegraphics[width=.45\textwidth]{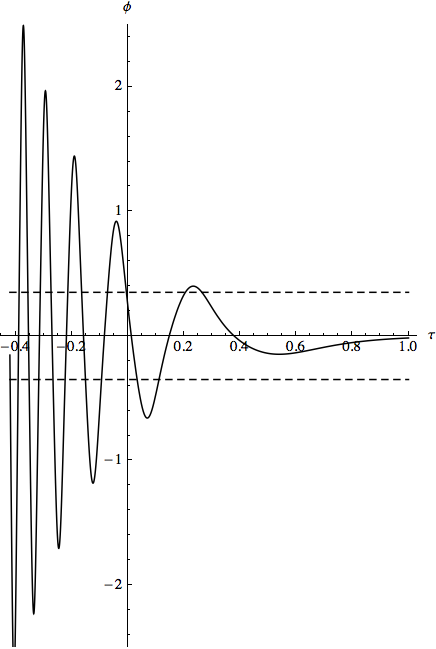}
\caption{This figure complements the description of the above figures, by fixing a point of the background space, and showing only the $\tau$-dependence of the field; due to the presence of the exponential factor $\exp(\frac{3}{2}m^2\tau)$, which is increasing as $\tau\rightarrow-\infty$, we have the asymptotic $lim_{\tau\rightarrow -\infty}\phi(x,\tau)=\pm\infty$; additionally we have $lim_{\tau\rightarrow +\infty}\phi(x,\tau)=0$. The edges and holes shown in the figures \ref{ffractal}, and \ref{fffractal} do not represent singularities, but represent the cuts represented by the dashed lines in this figure.}
 \label{ffffractal}
  \end{center}
 \end{figure}

A self-similar solution for the Eq. (\ref{one-dim}) can be found for SSB scenarios with $m^{2}<0$;
\begin{eqnarray}
     \phi (x,\tau ) \!\! & = & \!\! exp(\frac{3}{2} m^{2}\tau ) \sin \big( \frac{1}{2} \sqrt{2|m^{2}|} x + C_{1}\big) \psi (\xi ), \label{similar1} \\
     \xi \!\! & = & \!\!exp \big( \frac{3}{2} m^{2} \tau \big) \cos \big( \frac{1}{2} \sqrt{2|m^{2}|} x + C_{1}\big) + C_{2}, \label{similar2} \\
     \frac{d^{2}\psi}{\partial\xi^{2}} \!\! & = & \!\! 2 \frac{\lambda}{m^{2}} \psi^{3}, \label{similar3}
\end{eqnarray}
where the solution for the Eq. (\ref{similar3}) can be expressed in terms of an elliptic integral,
\begin{eqnarray}
     \xi \!\!  =  \!\! \pm \int \frac{d\psi}{\sqrt{C^{2}_{3} + \frac{\lambda}{m^{2}}\psi^{4}}} \label{elliptic} 
     \!\! =  \!\! \pm \frac{1}{\sqrt{C_{3}}}\sqrt{\frac {|m^{2}|} {\lambda}}F(i \sinh^{-1}(\sqrt{\frac{1}{C_{3}}\frac {|m^{2}|} {\lambda}}\psi)|-1) ,  \label{elliptic} 
\end{eqnarray}     
where, for simplicity we have assumed that $C_{3}>0$, for obtaining a real solution; $F$ corresponds to the elliptic function of the first kind. In order to put the function $\psi$ as a function of $\tau$, we consider that if $G=F(J|m)$, then the Jacobi Amplitude is the inverse of $F$, $J=Am(G|m)$, whose properties are well known. Therefore,
\begin{equation}
\psi(\xi) = \sqrt{C_{3}}\sin(Am[b \sqrt{C_{3}\frac{\lambda}{|m^2|}}\xi|-1]);
\label{ja}
\end{equation}
where $b=\pm$1; once the above expression is substituted into the Eq. (\ref{similar1}), we obtain the complete expression for the field $\phi(x,\tau)$; the figures (\ref{ffractal}), and
(\ref{fffractal}) illustrate {\it views from the top} for the $\phi(x,\tau)$-configurations for $b=1$, with the idea of having a complete visualization of the periodicity of the pattern in the spatial direction, and the self-similarity in the {\it backward} $\tau$-direction. The figure \ref{ffffractal} show the configuration $\phi(x_{f}, \tau)$, where $x_{f}$ is a fixed point of the flat background.

\begin{figure}[H]
  \begin{center}
    \includegraphics[width=.5\textwidth]{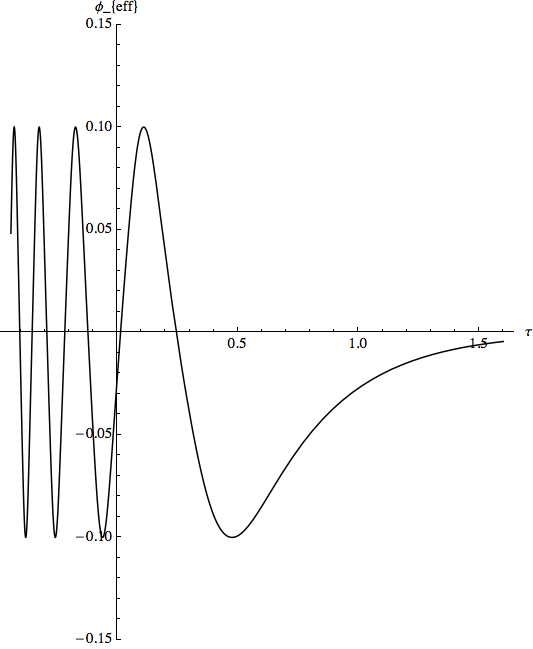}
\caption{This figure must be compared with its {\it dual} version shown in the figure \ref{ffffractal}, which is divergent in the domain $\tau<0$; in this case we have the asymptotic $lim_{\tau\rightarrow-\infty}\phi_{eff}=\pm \sqrt{C_{3}}$, and $lim_{\tau\rightarrow +\infty}\phi_{eff}=0$; these limits must be complemented with the asymptotic limits  for the pair $(m^2,\lambda)_{eff}$ described above.}    
\label{phi-effective}
  \end{center}
 \end{figure}

There exist no phase transitions of any order; the field $\phi$ and its derivatives are described in terms of well-behaved functions, including the Jacobi Amplitude and its derivatives. Specifically, the Jacobi Amplitude has no singularities or divergences; now,  all derivatives of this function can be described in a closed form in terms of other Jacobi functions, which are also well-behaved; namely, $\frac{\partial Am(z|m)}{\partial z}=dn(z|m)$, $\frac{\partial dn(z|m)}{\partial z}=-m sn(z|m)cn(z|m)$, $\frac{\partial sn(z|m)}{\partial z}= dn(z|m)cn(z|m)$, and $\frac{\partial cn(z|m)}{\partial z}= -sn(z|m)dn(z|m)$; hence, the three Jacobi functions $dn,sn,cn$, which are well-behaved in all interval, describe  completely all order derivatives of the dynamical field
$\phi$, without divergences and singularities.
 This description of the deformation of the theory based on a field $\phi$ increasing without limit as $\tau\rightarrow-\infty$, with  the pair $(m,\lambda )$ fixed, can be transformed into a {\it dual} version with a field $\phi$ bounded, and the pair $(m,\lambda)$ increasing without limit. Let us consider the potential, with the explicit form the field $\phi$ given in (\ref{similar1}),
 \begin{equation}
 V(\phi,m,\lambda)= \frac{1}{2}[m^2 \exp(3 m^{2}\tau) ][\sqrt{C_{3}}\sin \big( \frac{1}{2} \sqrt{2|m^{2}|} x + C_{1}\big) \psi (\xi )]^2
 +\frac{1}{4!} [\lambda \exp(6 m^{2}\tau)][\sqrt{C_{3}}\sin \big( \frac{1}{2} \sqrt{2|m^{2}|} x + C_{1}\big) \psi (\xi )]^4,  \label{similar-effective}\end{equation}
 where the global exponential factor has been adsorbed completely in the pair $(m,\lambda)$, which have now an {\it effective} $\tau$-dependence; note that the global factor does not affect the negativity of $m^2$, and the positivity of $\lambda$ (thus, there is no a qualitative change, and we have always SSB scenarios), with the asymptotic
 $lim_{\tau\rightarrow-\infty}(m^2,\lambda)_{eff}=+\infty$, corresponding to a massive and strongly self-interacting version, and  $lim_{\tau\rightarrow+\infty}(m^2,\lambda)_{eff}=0$, corresponding to a massless and  (asymptotically) free version. The scalar field has also an {\it effective} configuration given by $\phi_{eff}\equiv\sqrt{C_{3}}\sin \big( \frac{1}{2} \sqrt{2|m^{2}|} x + C_{1}\big) \psi (\xi )$, where the  $\tau$-dependence is through the 
 expressions (\ref{similar2}), and (\ref{ja}); therefore, by depending only on the $\sin$, and $\cos$ functions, the effective expression for the field is now bounded; the figure \ref{phi-effective} illustrates the behavior of the field for a fixed background point. From the perspective of this dual description, the solitonic character of this solution is more clear, since it maintains its profile as $\tau\rightarrow -\infty$,  self-maintaining the formation of the {\it branching pattern} described above.
Furthermore, the two lowest-energy states for the potential (\ref{similar-effective}) in this {\it dual} version,  are localized now at $[\phi_{eff}]_{vac}= \pm \sqrt{-\frac{m^2}{\lambda} }\exp{(-\frac{3}{2}m^2\tau)}$, with a hight given by $V_{vac}=-(m_{eff})^4/\lambda_{eff}=-m^4/\lambda$, which is constant under evolution. Therefore, maintaining fixed $V_{vac}$, we have the asymptotic $lim_{\tau\rightarrow-\infty}[\phi_{eff}]_{vac}=0$,  and the two minimums are closing; the meeting will take a infinite time.  Additionally, $lim_{\tau\rightarrow +\infty}[\phi_{eff}]_{vac}=\pm\infty$, and the minimums are moving away. In this dual version, the self-similarity patterns shown in the figures \ref{ffractal} and \ref{fffractal} are maintained, but with the {\it peaks} suppressed, according to figure \ref{phi-effective}. 
As opposed to the case of ``traveling waves", in this case the two-lowest energy states are stable under the $\tau$-deformation; hence,  the choice of a vacuum (and the spontaneous symmetry breakdown) is possible for all $\tau$.
 
  The case of solutions with a Jacobi Amplitude with complex argument will be developed elsewhere; this case necessarily will develop the {\it doubling} of the fields, and a more complicated self-similarity patterns formation. \\

  \noindent {\uno II.  Concluding remarks} 
  
The criterion that we have used for constructing the parameters and fields gradient flows is the steepest ascent and descent for a functional, which will be reached 
only when the complete infinite series be taken into the account, in particular that related with the $\phi$-gradient flow (\ref{inf3}). The results reported here, have been obtained considering only the lowest order term in that gradient flow; higher order corrections must be considered in analogy with a loop expansion in the usual renormalization group flows scheme. However, within the one-order treatment developed in this work, one can consider direct generalizations that deserve more explorations, and that may lead to nontrivial extensions, namely, the gradient flows (\ref{inf1}), and (\ref{inf2}) admit additional infinite series of quadratic powers of the dynamical field $\phi$, according to the gradient flow for the functional (\ref{var-2}); this coupling between the gradient flows will lead to a scheme where the field $\phi$ 
will play a key role in the phase transitions; works in this direction are in progress. Additionally, the Hamiltonian formulation, and the subsequent quantization of the  theory based on the present results are being developed, and will be reported in forthcoming works.

\vspace{1em}
\begin{center}
{\uno ACKNOWLEDGMENTS}
\end{center}
This work was supported by the Sistema Nacional de Investigadores (M\'{e}xico), and VIEP-BUAP(M\'{e}xico); the analysis of the differential equations, and the figures, were made using Mathematica.

\end{document}